# Characterization of Performance Anomalies in Hadoop

## THESIS

Presented in Partial Fulfillment of the Requirements for the Degree Master of Science in the Graduate School of The Ohio State University

By

Puja Gupta

Graduate Program in Computer Science and Engineering

The Ohio State University

2015

Master's Examination Committee:

Professor Christopher Stewart, Advisor

Professor Spyros Blanas

**Abstract**

With the huge variety of data and equally large-scale systems, there is not a unique execution setting for these systems which can guarantee the best performance for each query. In this project, we tried so study the impact of different execution settings on execution time of workloads by varying them one at a time. Using the data from these experiments, a decision tree was built where each internal node represents the execution parameter, each branch represents value chosen for the parameter and each leaf node represents a range for execution time in minutes. The attribute in the decision tree to split the dataset on is selected based on the maximum information gain or lowest entropy. Once the tree is trained with the training samples, this tree can be used to get approximate range for the expected execution time. When the actual execution time differs from this expected value, a performance anomaly can be detected. For a test dataset with 400 samples, 99% of samples had actual execution time in the range predicted time by the decision tree.

Also on analyzing the constructed tree, an idea about what configuration can give better performance for a given workload can be obtained. Initial experiments suggest that the impact an execution parameter can have on the target attribute (here execution time) can be related to the distance of that feature node from the root of the constructed decision tree. From initial results the percent change in the values of the target attribute for various value of the feature node which is closer to the root is 6 times larger than when that same feature node is away from the root node. This observation will depend on how well the decision tree was trained and may not be true for every case.

**Dedication**

This document is dedicated to my family and friends.

# Acknowledgments

I thank my advisor, Christopher Stewart, for being patient with me and supporting me in my mistakes. We did try out couple of ideas before working on this one but he was always supportive and easy to reach out to. Also giving encouragement for some things I might not have done.

I am thankful to all the extremely talented Professors at Ohio State University with whom I got a chance to interact for making my stay at OSU worthwhile. I would also like to express my gratitude to Professor Spyros Blanas for being on the committee chair.

The path through graduate school would have been difficult without constant support and help of current and former fellow buckeyes. Also it would have been impossible to stay so far away from home without all the friends in Columbus. Also a special mention of my best friends back in India who always bear with me in my best and worst times.

Finally, big thanks to mom and dad for always being my strength even though mom had been skeptical if I would stay long enough in a different country to complete my masters, my brother Bharat for being critical of me and bringing the confidence in me when I had none left at times.

**Vita**

2006 ............................................................ High School, Mount Carmel High School, Akola, India

2006-2010 ................................................... B.Tech. Computer Science & Engineering, Government College of Engineering, Pune, India

2010 ............................................................ Intern, Symantec, Pune, India

2010-2013 ................................................... Systems Engineer, TCS, Pune, India

2014 ............................................................ Intern, Qualcomm, San Diego, USA

2013-present .............................................. Graduate student/Graduate Teaching Assistant, CSE Department, Ohio State University

**Fields of Study**

Major Field:  Computer Science and Engineering

# Table of Contents

# List of Figures

# List of Tables

# Chapter 1: Introduction

Just like any other complex machine, Hadoop system also needs tuning to get the best performance. Although Hadoop distribution comes with default configuration it may not be the best for any given cluster. The optimal values for these configuration parameters depend on the application characteristics, size and content of input data and the cluster setup. To understand how configuration settings can affect the performance of the cluster, consider the execution time obtained for a query which was run with a replication factor of 3 for HDFS blocks and dataset size of 1GB. All other settings were kept constant across the experiments and we only change the number of nodes in the cluster.

| Number of nodes | 2 | 4 | 8 | 12 | 16 |
|---|---|---|---|---|---|
| Execution time(minutes) | 27.26 | 32 | 34.18 | 39.22 | 41.32 |

Table 1Effect of change in number of nodes on execution time

As seen from the table, the execution time was best with small cluster of size two and it kept degrading as we increased the size of the cluster. We would expect to get better performance with more number of machines but since the dataset size was not large enough, increasing the number of nodes split the job in more number of tasks and the overhead to manage those tasks affected the performance.

For my master's thesis, I identified some important execution settings that impact the performance of a job in Hadoop cluster and run benchmark tests by varying value of one parameter at a time. The values chosen for various configurable parameters in Hadoop cluster like replication factor, block size, number of nodes etc. when you start your job execution define the execution setting for that job. We used BigBench[2] which is Big data benchmark developed by Intel and Cloudera along with other industry partners. BigBench query workload consists of 30 queries, of which 10 queries operating on structured data is taken from TPC-DS workload and remaining 20 queries are adapted from McKinsey [17] report on big data use cases. Of these seven queries run on semi-structured part of schema and 6 queries on unstructured part. Semi-structured data consists of registered/guest user clicks on retailer website whereas unstructured data consists of product reviews submitted online.

We tested the benchmarks across about 50 execution settings for 30 different queries. Some of the queries could not complete execution so might not be included in study. We measured response time, throughput and CPU utilization under each setting. The tests captured complex interactions between the execution parameters. Decision trees provide a graphical representation of the interactions between parameters. A decision tree is an acyclic graph where nodes represent a range of setting for one parameter, e.g., number of nodes. A path from the root of the tree to the leaves represents a range of execution settings. Leaves represent the absolute performance achieved under those settings. Leaves may cover range of performances. This graphical model provides a hierarchy on the impact of various parameters and their interactions.

We made several critical decisions when we created our decision tree. First, we decided to use a range of performance for leaf nodes instead of absolute performance. We devised a K-means clustering approach based on the knee-points in the execution time CDF. Specifically, all the test samples are sorted based on their execution time and the set is divided into subsets after recognizing the knee point values in the execution time range. This is described in more detail in Figure 2. The target attribute for decision tree is a range of values marked by these knee point values.

Second, we used maximum information gain to build our decision tree. Information gain characterizes the reduction in chaos provided by knowing the setting of an execution parameter. Each split in our tree means that the selected parameter best splits the tested settings for homogeneous outputs. In other words, the constructed decision tree signifies that each execution setting which was used as feature in decision tree has different impact on execution time for different type of workloads. Once this decision tree is constructed, we select two branches of decision tree which split on similar attributes but at different levels in the decision tree. These two branches represent two kinds of workloads. We change the values of attributes for these branches and calculate the change in target values.

Initial results indicate the impact a feature attribute can have on the target attribute can be related to its distance from root node in the decision tree. The closer feature node is to the root node, the more significant effect it can have on target attribute. Figure 4 explains this behavior in detail.

The rest of this thesis is organized as follows. Chapter 2 discusses some information about the technologies used. Chapter 3 discusses the approach and experiments done. Chapter 4 discusses the results from the experiments. Chapter 5 talks about related work.

## Chapter 2: Background

### A. Hadoop Framework

Hadoop [1] is an open source software framework designed for distributed processing of large scale data across clusters of computers. The important components in Hadoop 2.0 are:

- *Hadoop distributed file system (HDFS):* A distributed file system that provides high-throughput access to application data.
- *Hadoop YARN*: A framework for job scheduling and cluster resource management.
- *Hadoop MapReduce*: A YARN-based system for parallel processing of large data sets.

Hadoop framework can scale up from single node to thousands of machines where each machine has its own storage and computing power. The different daemons running when you have Hadoop framework up and running are:

- *Name node:* Name node is part of master. Each file in HDFS is divided into blocks and stored on different data nodes. Name node is responsible to store mapping between blocks and data nodes.
- *Data node:* Data node is located on each slave machine. Data nodes store blocks of data in their local system and tells name node about blocks it has stored locally.

- *Resource manager:* Resource manager is part of master node. It manages resources and schedules applications atop of YARN.
- *Node manager:* Node manager is located on each slave. It receives map-reduce jobs and executes them locally and communicates to resource manager when job is done.

**B. BigBench**

BigBench [2, 6] is an end-to-end big data benchmark which consists of 30 queries with a mix of structured, unstructured and semi-structured data. It employs PDGF, a parallel data generator to produce large amounts of data in scalable fashion. Similar to many big data systems BigBench also employs batch processing. BigBench uses four open source software frameworks to implement the 30 queries: Apache Hadoop, Apache Hive, Apache Mahout and Natural Language Processing Toolkit.

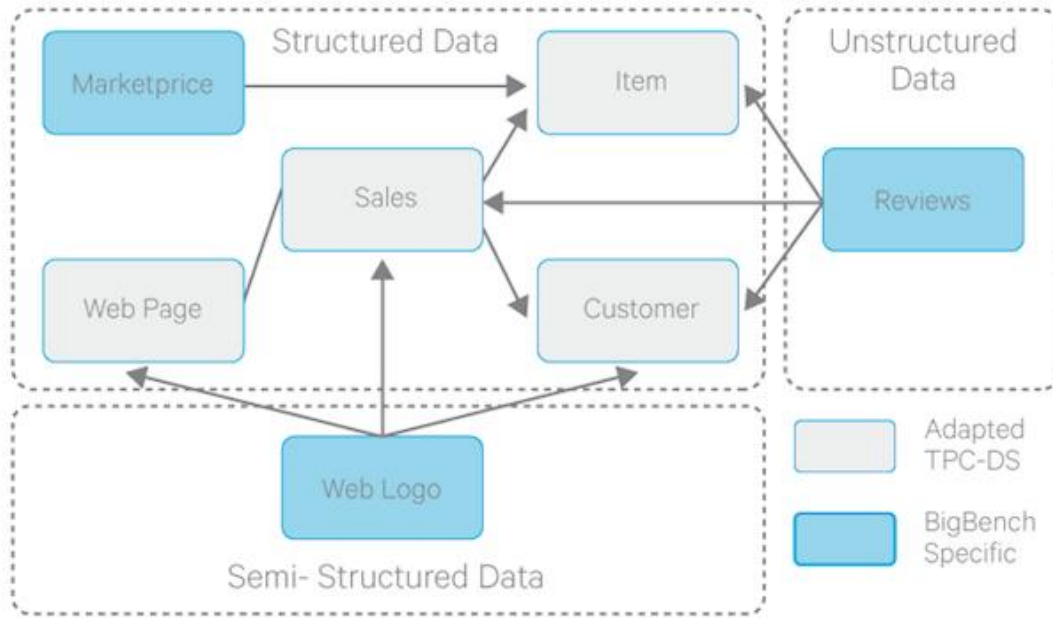

Figure 1Big Bench data model specification

Figure1 shows BigBench's schema for structured, unstructured and semi-structured data. BigBench is structured data intensive but it covers large variety of use cases.

### C. Decision tree

Decision tree [4] is a tree like structure where each internal node represents test on the attribute, each branch represents outcome of the test and each leaf node represents label i.e. a decision taken after evaluating all attributes. We used ID3 heuristic for feature selection purpose. It calculates the information gain for each of the feature, to select highly discriminate feature. When deciding the next attribute node, we choose the attribute which maximizes the information gain or the one with lowest entropy. In other words, best attribute is the one which is capable of discriminating samples that belong to different classes. To calculate information gain by splitting on a attribute, we calculate the frequency of each of the values of attribute in dataset. Following this, we calculate entropy for data set with new division of data derived by using the chosen attribute to classify the records in the dataset. Subtracting this new entropy from entropy of current dataset gives the information gain that can be achieved by choosing the attribute as next node. This calculation is done at every node to decide the next node and build sub-tree until all the attributes are used or when all the remaining ones give same values.

# Chapter 3: Methodology

For each benchmark trial, we store all the input and output data in Hadoop distributed file system [HDFS]. We used default settings of HDFS of three replicas per block without compression and 64 MB block size. We also tested other configurations like having single replica per block and up to 100% replication of blocks across the cluster. After each benchmark run finishes for particular node scaling level and replication factor, we delete data directories on each node and reformat HDFS so that next set of input data is replicated uniformly.

Some of the configuration parameters for Hadoop whose values directly affect performance of Hadoop cluster are:

- *Number of nodes*: Since each node was single core machine, maximum number of tasks was limited to number of nodes in cluster. We tested the benchmark on as less as 2 nodes up to 16 nodes in the cluster.

- *Size of dataset*: In BigBench, PDGF can generate data of any size between 1GB and 1 PB. We tested with sizes of 1 GB and 10GB for the data in benchmark tests.

- *Replication factor*: Each file is split into blocks based on block size set in Hadoop config file. Replication factor controls number of copies of each block that is maintained on different nodes in the cluster. Replication factor greater than number of data nodes in the cluster does not make much sense. Hence it can have any value between 1 and number of nodes in the cluster. We tested for values of replication from 1 to number of nodes in cluster (100%) replication. However with bigger cluster, the chances of any node going down was higher and hence getting benchmark results with replication as 1 was difficult.

- *Collocation*: Hadoop has couple of daemons running each for HDFS and YARN components which was described in section II. We tested performance by having data node and node manager running on separate machines instead of on same slave machines. Similarly for master node, resource manager and name node were on different machines.
- *Block size*: HDFS will split input data into smaller blocks of block size and these blocks are used as input for map tasks. The block size will determine the number of tasks that are launched for a job. Thus block size seems to be dependent on size of the cluster because increasing block size does not necessarily improve performance for each cluster configuration. For smaller cluster, larger block size will ensure lesser number of map tasks created and there will be less overhead. This can result in better performance.
- *Master with data node*: We tested having a slave (data node) running along with master node and not having data node running on master node. For some benchmark queries having a slave running along with master suggests better performance but for most having them separate gives better performance.

Execution time for different queries is collected by varying the above said configurations and percentage gain is calculated on the sorted list of these values. For each of the benchmark test, all the data from data node is cleared and the name node is formatted so that replication of blocks is consistent. The execution time is divided into range of values based on knee point values of percentage gain. This range of values was used as target attribute for the decision tree which is described next.

For supervised learning algorithm like decision tree, we need to have labeled samples as training samples. Using this it selects feature that can discriminate samples that belong to different classes. The training sample used in our study has six attributes namely query_number, replication_factor, dataset_size, number_of_nodes, block_size, collocation and a target label which indicates a range for execution time. In case there is more than one possible value for target attribute, all possible values along with their probability is listed. This can be the case because not all the features might be used in the particular branch of decision tree and hence not being unique there can be more than one possible value for target attribute. Query number is used as an attribute because it identifies with the type of operation being performed. Each query in BigBench is related to specific operations like Logistic regression, Bayesian Classifier, inner join, group by etc and query number can help to relate which feature is more significant for that kind of operation.

Once the decision tree is trained using training samples, we can get an estimated range for the execution time by providing values for the feature attributes. Using this estimated range, performance anomaly can be detected. Query which takes longer execution time than the predicted range can be recognized and analyzed to improve their performance. The accuracy for estimated target attribute will depend on how well the decision tree was trained for that kind of operation. The more training samples for that particular operation will guarantee more accurate results.

## Chapter 4: Experimental Setup and Results

The test environment for our experiments consisted of up to 20 virtual machines connected together in a cluster. Each virtual machine was 2GHz single processors with 4GB RAM and 64 bit Linux CentOS.

Benchmark queries were run on this cluster for each different configuration setting and the execution time for that run was recorded. Then all test samples which consist of configuration parameter as attribute and execution time as target were sorted based on execution time. The knee points in this distribution were identified to divide execution time range into different smaller subsets.

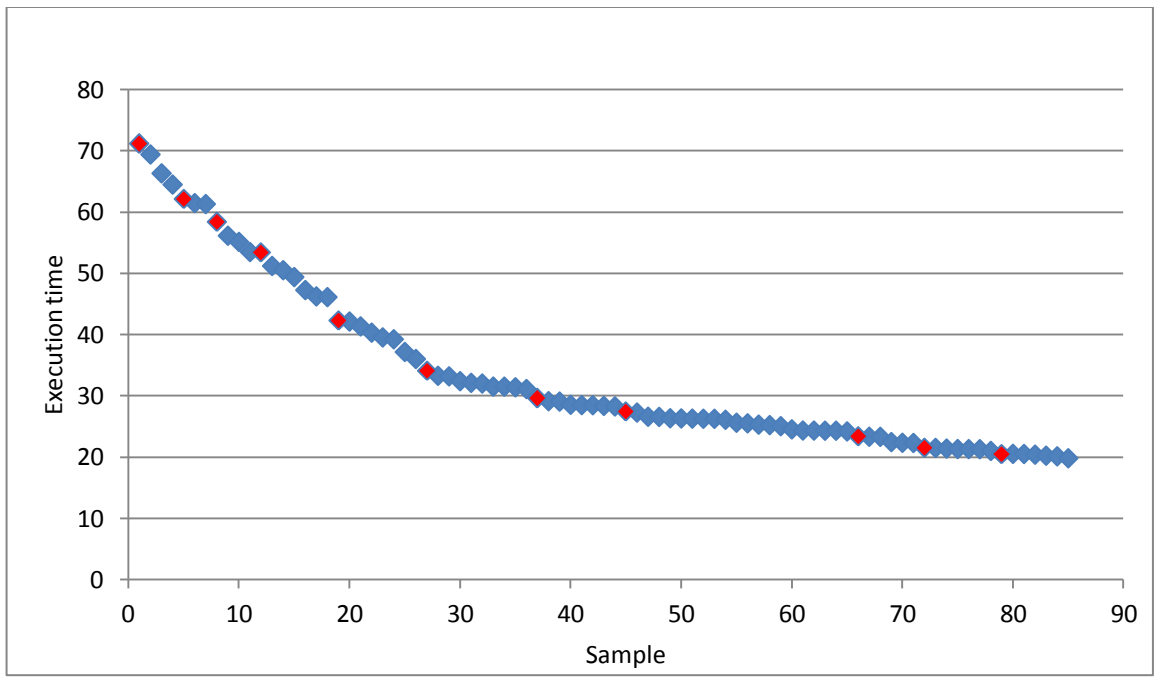

Figure 2 Detecting knee points in execution time cdf

In the graph above, consider the distribution of execution times for 85 sample tests arranged in decreasing order of execution time. The knee points are highlighted in the distribution and these knee points determine range denoted by leaf nodes of decision tree. Consider part of decision tree for queries 4 and 5 shown below. Entire decision tree for all 30 queries in BigBench can be found in Appendix A. The left most node 'query' represents root of the decision tree. Each key-value pair in the decision tree below represents execution parameter and value chosen for that parameter. The leaf nodes to the right of the tree represent range for execution time in minutes along with probability distribution for that range. If more than one range is present for a leaf node, we can choose the best value with highest probability. As shown in figure, branch for query 4 next splits on replication factor and then on number of nodes. Also branch for query 5 splits on number of nodes and then on replication factor. The root of the decision tree is query number which helps to identify the type of operation being performed.

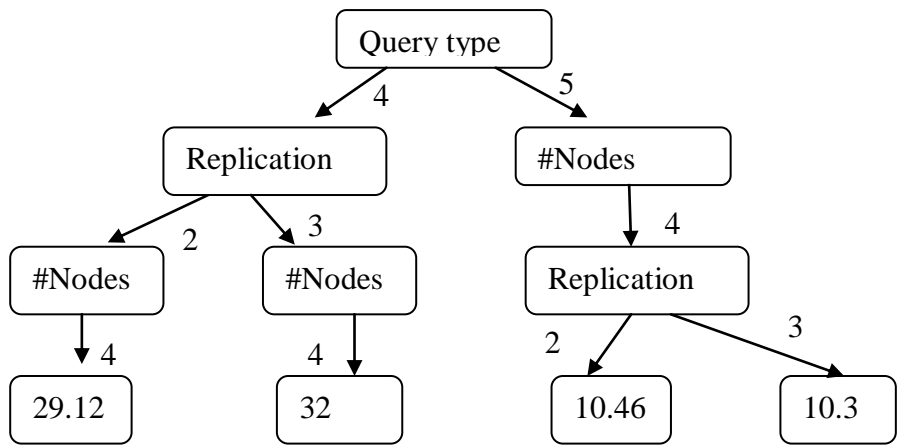

Figure 3 Effect of level in decision tree on execution time

The above diagram represents only part from entire decision tree so as to make the following observation understand better. Consider two workloads for query 4 and query 5. For the first experiment shown number of nodes for both workloads was 4, dataset size was 1GB and we vary replication factor for both workloads from 2 to 3. As seen in decision tree above, for query 4 branch, replication factor was at level 2 in the constructed decision tree whereas in the same decision for query 5 branch, replication factor was at level 3. The percent change in execution time for query 4 which was closer to root node is ~9.8% on changing replication factor whereas for query 5 the percent change in execution time is ~1.5%.

Now again consider the same workloads as in previous experiment with replication factor set to 3, dataset size of 1GB and we vary number of nodes from 8 to 16. As seen in decision tree number of nodes was at level 3 for query 4 whereas at level 2 for query 5. For query 4 which was further from root, percent change in execution time on changing number of nodes was ~20.88% whereas for query 5 which was closer to root node was ~37.33%.

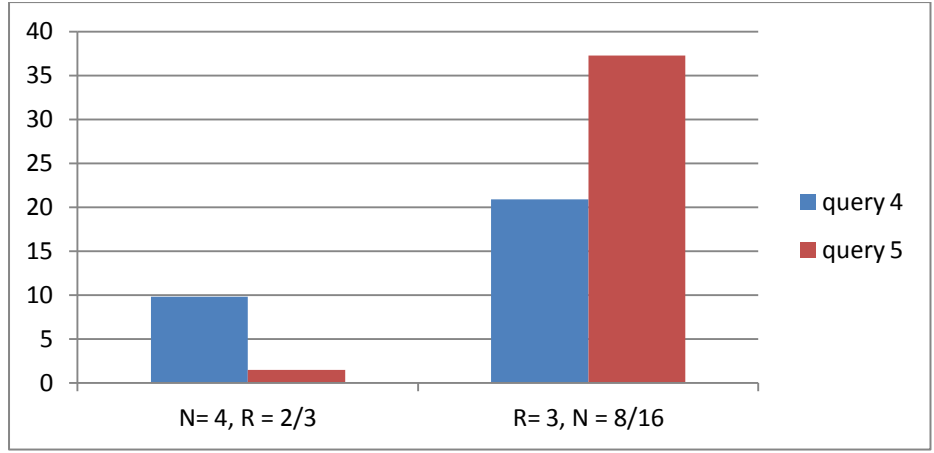

Figure 4 Percent change in execution time

The constructed decision tree was used to predict execution time for test samples. Using entire test dataset of ~400 samples, it was found that almost 99% of queries got their execution time in the range predicted using trained decision tree.

# Chapter 5: Related work

There are several works that have focused on Hadoop/MapReduce performance tuning. Herodotos et al. [10-12] proposed a cost based optimization technique to help users identify good job configurations for MapReduce applications. They used a profiler to get concise statistics like cost estimation, a what-if engine to reason about the impact of parameter configuration settings and a cost-based optimizer to find good configurations. Our approach is different from this work as we get execution times from test runs and use supervised decision tree to get estimated range for execution time of job. Some simulation based performance tuning techniques have also been explored [13-14]. These systems do not actually change configuration parameters but provide estimate of application performance using their designed system. Ali R. Butt et al. [15] proposed a dynamic performance tuning system which monitors job execution and tunes associated parameters. However this approach has some limitations as some parameters cannot be changed at runtime to have immediate effect on already executing job. Rather than tuning performance and resource usage tradeoff, Kelley et al. [19] allow the answer quality to degrade to save resources. On Map-Reduce style workloads; they observe that the quality degradation is non-linear with regard to resource savings. However, the work does not currently consider a wide range of parameters that may affect this trade off. Also job monitoring and collecting statistics adds an overhead to overall execution time of the job.

Stewart et al. [18] proposed a framework for depicting performance anomaly manifestations using decision tress for classification.

# Chapter 6: Conclusion

A simple supervised algorithm like decision tree can be used to get the expected time range for queries. If the actual execution value does not match this expected time range, a performance anomaly can be detected. By analyzing decision tree for that particular workload, an idea about what kind of execution setting can give better performance can be obtained. Initial results show that for test dataset with around 460 samples almost 95% of the values match with the expected values. Also initial results suggest that the closer the feature node to the root node, the more significant impact it can have on the target attribute. This observation might not be true for every feature in every workload but it appears promising since node closer to root has more information gain and hence the range for target attribute of the branches from this node is more distributed. Also the accuracy of results will depend on how well the decision tree is trained using sample data.

# Appendix A: Decision tree for BigBench queries

The following text represents decision tree where root node is on the leftmost side(in this case it is query). Parse the tree left to right where rightmost nodes represent leaf of the tree with a range for execution time and a probability associated for that estimate.

```
{ 'query': { 1: { 'nodes': { 2: { 'replication': { 2: <ProbDist 8.6-8.3=1.0>,
                                                    3: { 'blk_range': { 64: <ProbDist 6.7-5.8=1.0>,
                                                                        128: <ProbDist 5.8-4.54=1.0>}}}},
                             3: <ProbDist 7.7-6.7=1.0>,
                             4: { 'replication': { 1: <ProbDist 8.6-8.3=1.0>,
                                                   2: <ProbDist 8.6-8.3=1.0>,
                                                   3: <ProbDist 6.7-5.8=1.0>,
                                                   4: <ProbDist 10.27-9.59=1.0>}},
                             5: <ProbDist 7.7-6.7=1.0>,
                             6: <ProbDist 7.7-6.7=1.0>,
                             8: { 'replication': { 1: <ProbDist 6.7-5.8=1.0>,
                                                   2: <ProbDist 10.6-10.27=1.0>,
                                                   3: { 'blk_range': { 64: <ProbDist 8.6-8.3=1.0>,
                                                                       128: <ProbDist 9.59-8.6=1.0>}}}},
                             12: { 'replication': { 1: <ProbDist 11.59-11.04=1.0>,
                                                    2: <ProbDist 8.15-7.7=1.0>,
                                                    3: { 'data_size': { 1: <ProbDist 9.59-8.6=1.0>,
                                                                        10: <ProbDist 10.6-10.27=1.0>}},
                                                    6: <ProbDist 7.7-6.7=1.0>}},
                             16: { 'replication': { 2: <ProbDist 12.58-11.59=1.0>,
                                                    3: <ProbDist 8.6-8.3=1.0>,
                                                    4: <ProbDist 9.59-8.6=1.0>}}}},
             2: { 'replication': { 1: { 'nodes': { 4: { 'colocated': { 0: <ProbDist 10.27-9.59=1.0>,
                                                                       1: <ProbDist 8.6-8.3=1.0>}},
                                                   8: <ProbDist 9.59-8.6=1.0>,
                                                   12: <ProbDist 16.58-15.8=1.0>}},
                                   2: { 'nodes': { 2: <ProbDist 12.58-11.59=1.0>,
                                                   4: <ProbDist 10.27-9.59=1.0>,
                                                   8: <ProbDist 12.58-11.59=1.0>,
                                                   12: <ProbDist 11.59-11.04=1.0>,
                                                   16: <ProbDist 12.58-11.59=1.0>}},
                                   3: { 'nodes': { 2: { 'blk_range': { 64: <ProbDist 8.6-8.3=1.0>,
                                                                       128: <ProbDist 7.7-6.7=1.0>}},
                                                   3: <ProbDist 10.6-10.27=1.0>,
                                                   4: <ProbDist 8.6-8.3=1.0>,
                                                   5: <ProbDist 9.59-8.6=1.0>,
                                                   6: <ProbDist 10.27-9.59=1.0>,
                                                   8: { 'blk_range': { 64: <ProbDist 11.59-11.04=1.0>,
                                                                       128: <ProbDist 13.9-12.58=1.0>}},
                                                   12: { 'data_size': { 1: <ProbDist 14.56-13.9=1.0>,
                                                                        10: <ProbDist 13.9-12.58=1.0>}},
                                                   16: <ProbDist 12.58-11.59=1.0>}},
```

```
                              4: <ProbDist 11.59-11.04=1.0>,
                              6: <ProbDist 11.04-10.6=1.0>}},
       3: {  'colocated': {   0: <ProbDist 7.7-6.7=1.0>,
                    1: {  'nodes': {  2: {  'replication': {   2: <ProbDist 7.7-6.7=1.0>,
                                                         3: <ProbDist 5.8-4.54=1.0>}},
                                      3: <ProbDist 6.7-5.8=1.0>,
                                      4: {  'replication': {   1: <ProbDist 6.7-5.8=1.0>,
                                                         2: <ProbDist 7.7-6.7=1.0>,
                                                         3: <ProbDist 6.7-5.8=1.0>,
                                                         4: <ProbDist 7.7-6.7=1.0>}},
                                      5: <ProbDist 6.7-5.8=1.0>,
                                      6: <ProbDist 7.7-6.7=1.0>,
                                      8: {  'replication': {   1: <ProbDist 5.8-4.54=1.0>,
                                                         2: <ProbDist 8.6-8.3=1.0>,
                                                         3: <ProbDist 7.7-6.7=1.0>}},
                                      12: {  'replication': {   2: <ProbDist 7.7-6.7=1.0>,
                                                          3: {  'data_size': {   1: <ProbDist 9.59-8.6=1.0>,
                                                                         10: <ProbDist 13.9-12.58=1.0>}},
                                                          6: <ProbDist 7.7-6.7=1.0>}},
                                      16: {  'replication': {   2: <ProbDist 9.59-8.6=1.0>,
                                                          3: <ProbDist 7.7-6.7=1.0>,
                                                          4: <ProbDist 9.59-8.6=1.0>}}}}}},
       4: {  'replication': {   1: <ProbDist 37.19-30.38=1.0>,
                    2: {  'nodes': {  2: <ProbDist 56.12-42.27=1.0>,
                                      4: <ProbDist 30.38-26.58=1.0>,
                                      8: <ProbDist 42.27-37.19=1.0>,
                                      12: <ProbDist 71.12-58.39=1.0>,
                                      16: <ProbDist 42.27-37.19=1.0>}},
                    3: {  'nodes': {  2: {  'blk_range': {   64: <ProbDist 30.38-26.58=1.0>,
                                                       128: <ProbDist 26.58-23.39=1.0>}},
                                      3: <ProbDist 37.19-30.38=1.0>,
                                      4: <ProbDist 37.19-30.38=1.0>,
                                      5: <ProbDist 30.38-26.58=1.0>,
                                      6: <ProbDist 30.38-26.58=1.0>,
                                      8: <ProbDist 37.19-30.38=1.0>,
                                      12: <ProbDist 42.27-37.19=1.0>,
                                      16: <ProbDist 42.27-37.19=1.0>}},
                    4: <ProbDist 37.19-30.38=1.0>,
                    6: <ProbDist 37.19-30.38=1.0>}},
       5: {  'nodes': {  2: {  'replication': {   2: <ProbDist 15.8-14.56=1.0>,
                                            3: <ProbDist 9.59-8.6=1.0>}},
                    3: <ProbDist 11.59-11.04=1.0>,
                    4: {  'replication': {   1: {  'colocated': {   0: <ProbDist 15.8-14.56=1.0>,
                                                             1: <ProbDist 14.56-13.9=1.0>}},
                                       2: <ProbDist 10.6-10.27=1.0>,
                                       3: <ProbDist 10.6-10.27=1.0>}},
                    5: <ProbDist 11.59-11.04=1.0>,
                    6: <ProbDist 11.59-11.04=1.0>,
                    8: {  'replication': {   1: <ProbDist 13.9-12.58=1.0>,
                                       2: <ProbDist 14.56-13.9=1.0>,
                                       3: <ProbDist 11.59-11.04=1.0>}},
                    12: {  'replication': {   2: <ProbDist 15.8-14.56=1.0>,
                                        3: {  'data_size': {   1: <ProbDist 15.8-14.56=1.0>,
```

```
                                            10: <ProbDist 42.27-37.19=1.0>}},
                          6: <ProbDist 11.59-11.04=1.0>}},
              16: {   'replication': {   2: <ProbDist 16.58-15.8=1.0>,
                          3: <ProbDist 14.56-13.9=1.0>,
                          4: <ProbDist 12.58-11.59=1.0>}}}},
    6: {   'nodes': {   2: {   'replication': {   2: <ProbDist 18.56-17.58=1.0>,
                          3: {   'blk_range': {   64: <ProbDist 13.9-12.58=1.0>,
                                    128: <ProbDist 11.59-11.04=1.0>}}}},
              3: <ProbDist 15.8-14.56=1.0>,
              4: {   'replication': {   1: {   'colocated': {   0: <ProbDist 18.56-17.58=1.0>,
                                            1: <ProbDist 17.58-16.58=1.0>}},
                          2: <ProbDist 14.56-13.9=1.0>,
                          3: <ProbDist 13.9-12.58=1.0>}},
              5: <ProbDist 15.8-14.56=1.0>,
              6: <ProbDist 16.58-15.8=1.0>,
              8: <ProbDist 17.58-16.58=1.0>,
              12: {   'replication': {   2: <ProbDist 18.56-17.58=1.0>,
                          3: {   'data_size': {   1: <ProbDist 19.57-18.56=1.0>,
                                    10: <ProbDist 30.38-26.58=1.0>}},
                          6: <ProbDist 15.8-14.56=1.0>}},
              16: {   'replication': {   2: <ProbDist 20.53-19.57=1.0>,
                          3: <ProbDist 20.53-19.57=1.0>,
                          4: <ProbDist 19.57-18.56=1.0>}}}},
    7: {   'nodes': {   2: {   'replication': {   2: <ProbDist 23.39-21.5=1.0>,
                          3: {   'blk_range': {   64: <ProbDist 17.58-16.58=1.0>,
                                    128: <ProbDist 15.8-14.56=1.0>}}}},
              3: <ProbDist 20.53-19.57=1.0>,
              4: {   'replication': {   1: {   'colocated': {   0: <ProbDist 21.5-20.53=1.0>,
                                            1: <ProbDist 19.57-18.56=1.0>}},
                          2: <ProbDist 17.58-16.58=1.0>,
                          3: <ProbDist 16.58-15.8=1.0>}},
              5: <ProbDist 19.57-18.56=1.0>,
              6: <ProbDist 20.53-19.57=1.0>,
              8: {   'replication': {   1: <ProbDist 19.57-18.56=1.0>,
                          2: <ProbDist 23.39-21.5=1.0>,
                          3: <ProbDist 23.39-21.5=1.0>}},
              12: {   'replication': {   2: <ProbDist 26.58-23.39=1.0>,
                          3: {   'data_size': {   1: <ProbDist 26.58-23.39=1.0>,
                                    10: <ProbDist 37.19-30.38=1.0>}},
                          6: <ProbDist 20.53-19.57=1.0>}},
              16: {   'replication': {   2: <ProbDist 26.58-23.39=1.0>,
                          3: <ProbDist 30.38-26.58=1.0>,
                          4: <ProbDist 21.5-20.53=1.0>}}}},
    8: {   'nodes': {   2: {   'replication': {   2: <ProbDist 18.56-17.58=1.0>,
                          3: {   'blk_range': {   64: <ProbDist 13.9-12.58=1.0>,
                                    128: <ProbDist 11.59-11.04=1.0>}}}},
              3: <ProbDist 15.8-14.56=1.0>,
              4: {   'replication': {   1: {   'colocated': {   0: <ProbDist 18.56-17.58=1.0>,
                                            1: <ProbDist 16.58-15.8=1.0>}},
                          2: <ProbDist 13.9-12.58=1.0>,
                          3: <ProbDist 13.9-12.58=1.0>}},
              5: <ProbDist 15.8-14.56=1.0>,
              6: <ProbDist 14.56-13.9=1.0>,
```

```
          8: {   'replication': {   1: <ProbDist 15.8-14.56=1.0>,
                                   2: <ProbDist 18.56-17.58=1.0>,
                                   3: {   'blk_range': {   64: <ProbDist 17.58-16.58=1.0>,
                                                           128: <ProbDist 16.58-15.8=1.0>}}}},
         12: {   'replication': {   2: <ProbDist 19.57-18.56=1.0>,
                                   3: {   'data_size': {   1: <ProbDist 19.57-18.56=1.0>,
                                                           10: <ProbDist 26.58-23.39=1.0>}},
                                   6: <ProbDist 16.58-15.8=1.0>}},
         16: {   'replication': {   2: <ProbDist 26.58-23.39=1.0>,
                                   3: <ProbDist 21.5-20.53=1.0>,
                                   4: <ProbDist 18.56-17.58=1.0>}}}},
 9: {   'nodes': {   2: {   'replication': {   2: <ProbDist 11.59-11.04=1.0>,
                                              3: {   'blk_range': {   64: <ProbDist 7.7-6.7=1.0>,
                                                                      128: <ProbDist 5.8-4.54=1.0>}}}},
                     3: <ProbDist 6.7-5.8=1.0>,
                     4: {   'replication': {   1: {   'colocated': {   0: <ProbDist 9.59-8.6=1.0>,
                                                                       1: <ProbDist 8.15-7.7=1.0>}},
                                              2: <ProbDist 7.7-6.7=1.0>,
                                              3: <ProbDist 9.59-8.6=1.0>}},
                     5: <ProbDist 7.7-6.7=1.0>,
                     6: <ProbDist 7.7-6.7=1.0>,
                     8: {   'replication': {   2: <ProbDist 9.59-8.6=1.0>,
                                              3: {   'blk_range': {   64: <ProbDist 7.7-6.7=1.0>,
                                                                      128: <ProbDist 8.6-8.3=1.0>}}}},
                    12: {   'replication': {   1: <ProbDist 10.6-10.27=1.0>,
                                              2: <ProbDist 9.59-8.6=1.0>,
                                              3: {   'data_size': {   1: <ProbDist 10.6-10.27=1.0>,
                                                                      10: <ProbDist 19.57-18.56=1.0>}},
                                              6: <ProbDist 8.15-7.7=1.0>}},
                    16: {   'replication': {   2: <ProbDist 8.3-8.15=1.0>,
                                              3: <ProbDist 10.27-9.59=1.0>,
                                              4: <ProbDist 8.6-8.3=1.0>}}}},
10: {   'colocated': {   0: <ProbDist 9.59-8.6=1.0>,
                         1: {   'nodes': {   2: {   'replication': {   2: <ProbDist 9.59-8.6=1.0>,
                                                                       3: <ProbDist 5.8-4.54=1.0>}},
                                             3: <ProbDist 5.8-4.54=1.0>,
                                             4: <ProbDist 5.8-4.54=1.0>,
                                             5: <ProbDist 5.8-4.54=1.0>,
                                             6: <ProbDist 5.8-4.54=1.0>,
                                             8: {   'replication': {   2: <ProbDist 7.7-6.7=1.0>,
                                                                       3: {   'blk_range': {   64: <ProbDist 9.59-8.6=1.0>,
                                                                                               128: <ProbDist 7.7-6.7=1.0>}}}},
                                            12: {   'replication': {   1: <ProbDist 12.58-11.59=1.0>,
                                                                       2: <ProbDist 9.59-8.6=1.0>,
                                                                       3: {   'data_size': {   1: <ProbDist 6.7-5.8=1.0>,
                                                                                               10: <ProbDist 26.58-23.39=1.0>}},
                                                                       6: <ProbDist 9.59-8.6=1.0>}},
                                            16: {   'replication': {   2: <ProbDist 9.59-8.6=1.0>,
                                                                       3: <ProbDist 10.6-10.27=1.0>,
                                                                       4: <ProbDist 5.8-4.54=1.0>}}}}}},
11: {   'nodes': {   2: {   'replication': {   2: <ProbDist 10.6-10.27=1.0>,
                                              3: {   'blk_range': {   64: <ProbDist 7.7-6.7=1.0>,
                                                                      128: <ProbDist 6.7-5.8=1.0>}}}},
```

```
                      3: <ProbDist 8.3-8.15=1.0>,
                      4: {   'replication': {   1: {   'colocated': {   0: <ProbDist 10.6-10.27=1.0>,
                                                                        1: <ProbDist 9.59-8.6=1.0>}},
                                                2: <ProbDist 8.15-7.7=1.0>,
                                                3: <ProbDist 7.7-6.7=1.0>}},
                      5: <ProbDist 8.6-8.3=1.0>,
                      6: <ProbDist 9.59-8.6=1.0>,
                      8: {   'replication': {   2: <ProbDist 10.27-9.59=1.0>,
                                                3: <ProbDist 9.59-8.6=1.0>}},
                      12: {   'replication': {   2: <ProbDist 11.59-11.04=1.0>,
                                                 3: {   'data_size': {   1: <ProbDist 9.59-8.6=1.0>,
                                                                         10: <ProbDist 14.56-13.9=1.0>}},
                                                 6: <ProbDist 8.6-8.3=1.0>}},
                      16: {   'replication': {   2: <ProbDist 13.9-12.58=1.0>,
                                                 3: <ProbDist 10.6-10.27=1.0>,
                                                 4: <ProbDist 10.6-10.27=1.0>}}}},
          12: {   'nodes': {   2: {   'replication': {   2: <ProbDist 11.59-11.04=1.0>,
                                                         3: <ProbDist 7.7-6.7=1.0>}},
                      3: <ProbDist 9.59-8.6=1.0>,
                      4: {   'replication': {   1: {   'colocated': {   0: <ProbDist 10.6-10.27=1.0>,
                                                                        1: <ProbDist 9.59-8.6=1.0>}},
                                                2: <ProbDist 8.3-8.15=1.0>,
                                                3: <ProbDist 8.3-8.15=1.0>}},
                      5: <ProbDist 8.6-8.3=1.0>,
                      6: <ProbDist 9.59-8.6=1.0>,
                      8: {   'replication': {   2: <ProbDist 11.59-11.04=1.0>,
                                                3: {   'blk_range': {   64: <ProbDist 10.6-10.27=1.0>,
                                                                        128: <ProbDist 9.59-8.6=1.0>}}}},
                      12: {   'replication': {   2: <ProbDist 10.6-10.27=1.0>,
                                                 3: {   'data_size': {   1: <ProbDist 12.58-11.59=1.0>,
                                                                         10: <ProbDist 16.58-15.8=1.0>}},
                                                 6: <ProbDist 9.59-8.6=1.0>}},
                      16: {   'replication': {   2: <ProbDist 9.59-8.6=1.0>,
                                                 3: <ProbDist 11.59-11.04=1.0>,
                                                 4: <ProbDist 10.6-10.27=1.0>}}}},
          13: {   'nodes': {   2: {   'replication': {   2: <ProbDist 12.58-11.59=1.0>,
                                                         3: {   'blk_range': {   64: <ProbDist 8.6-8.3=1.0>,
                                                                                 128: <ProbDist 7.7-6.7=1.0>}}}},
                      3: <ProbDist 9.59-8.6=1.0>,
                      4: {   'replication': {   1: {   'colocated': {   0: <ProbDist 11.59-11.04=1.0>,
                                                                        1: <ProbDist 10.27-9.59=1.0>}},
                                                2: <ProbDist 9.59-8.6=1.0>,
                                                3: <ProbDist 9.59-8.6=1.0>}},
                      5: <ProbDist 9.59-8.6=1.0>,
                      6: <ProbDist 9.59-8.6=1.0>,
                      8: {   'replication': {   2: <ProbDist 12.58-11.59=1.0>,
                                                3: {   'blk_range': {   64: <ProbDist 11.59-11.04=1.0>,
                                                                        128: <ProbDist 10.6-10.27=1.0>}}}},
                      12: {   'replication': {   2: <ProbDist 13.9-12.58=1.0>,
                                                 3: {   'data_size': {   1: <ProbDist 13.9-12.58=1.0>,
                                                                         10: <ProbDist 30.38-26.58=1.0>}},
                                                 6: <ProbDist 11.59-11.04=1.0>}},
                      16: {   'replication': {   2: <ProbDist 12.58-11.59=1.0>,
```

```
                                    3: <ProbDist 12.58-11.59=1.0>,
                                    4: <ProbDist 11.59-11.04=1.0>}}}},
          14: {   'data_size': {   1: {   'colocated': {   0: <ProbDist 10.6-10.27=1.0>,
                                    1: {   'replication': {   1: <ProbDist 9.59-8.6=1.0>,
                                                              2: {   'nodes': {   2: <ProbDist 10.6-10.27=1.0>,
                                                                                  4: <ProbDist 7.7-6.7=1.0>,
                                                                                  8: <ProbDist 10.6-10.27=1.0>,
                                                                                  12: <ProbDist 9.59-8.6=1.0>,
                                                                                  16: <ProbDist 10.6-10.27=1.0>}},
                                                              3: {   'nodes': {   2: <ProbDist 6.7-5.8=0.5, 7.7-
6.7=0.5>,
                                                                                  3: <ProbDist 8.15-7.7=1.0>,
                                                                                  4: <ProbDist 7.7-6.7=1.0>,
                                                                                  5: <ProbDist 9.59-8.6=1.0>,
                                                                                  6: <ProbDist 9.59-8.6=1.0>,
                                                                                  8: <ProbDist 9.59-8.6=1.0>,
                                                                                  12: <ProbDist 11.59-11.04=1.0>,
                                                                                  16: <ProbDist 10.27-9.59=1.0>}},
                                                              4: <ProbDist 9.59-8.6=1.0>,
                                                              6: <ProbDist 9.59-8.6=1.0>}}}},
                             10: <ProbDist 14.56-13.9=1.0>}},
          15: {   'replication': {   1: <ProbDist 3.6-3.3=1.0>,
                                    2: {   'nodes': {   2: <ProbDist 3.6-3.3=1.0>,
                                                       4: <ProbDist 3.3-2.54=1.0>,
                                                       8: <ProbDist 4.54-3.9=1.0>,
                                                       12: <ProbDist 3.3-2.54=1.0>,
                                                       16: <ProbDist 3.6-3.3=1.0>}},
                                    3: {   'nodes': {   2: <ProbDist 2.54-1.52=1.0>,
                                                       3: <ProbDist 3.3-2.54=1.0>,
                                                       4: <ProbDist 2.54-1.52=1.0>,
                                                       5: <ProbDist 3.3-2.54=1.0>,
                                                       6: <ProbDist 2.54-1.52=1.0>,
                                                       8: {   'blk_range': {   64: <ProbDist 3.3-2.54=1.0>,
                                                                              128: <ProbDist 3.6-3.3=1.0>}},
                                                       12: {   'data_size': {   1: <ProbDist 4.54-3.9=1.0>,
                                                                               10: <ProbDist 5.8-4.54=1.0>}},
                                                       16: <ProbDist 4.54-3.9=1.0>}},
                                    4: <ProbDist 3.3-2.54=1.0>,
                                    6: <ProbDist 3.3-2.54=1.0>}},
          16: {   'replication': {   1: <ProbDist 4.54-3.9=1.0>,
                                    2: {   'nodes': {   2: <ProbDist 4.54-3.9=1.0>,
                                                       4: <ProbDist 3.3-2.54=1.0>,
                                                       8: <ProbDist 4.54-3.9=1.0>,
                                                       12: <ProbDist 3.6-3.3=1.0>,
                                                       16: <ProbDist 3.3-2.54=1.0>}},
                                    3: {   'nodes': {   2: <ProbDist 2.54-1.52=1.0>,
                                                       3: <ProbDist 3.6-3.3=1.0>,
                                                       4: <ProbDist 3.9-3.6=1.0>,
                                                       5: <ProbDist 3.3-2.54=1.0>,
                                                       6: <ProbDist 3.6-3.3=1.0>,
                                                       8: {   'blk_range': {   64: <ProbDist 3.3-2.54=1.0>,
                                                                              128: <ProbDist 3.6-3.3=1.0>}},
                                                       12: {   'data_size': {   1: <ProbDist 4.54-3.9=1.0>,
```
27

```
                                                10: <ProbDist 10.6-10.27=1.0>}},
                                  16: <ProbDist 5.8-4.54=1.0>}},
                    4: <ProbDist 3.6-3.3=1.0>,
                    6: <ProbDist 3.6-3.3=1.0>}},
     17: {  'nodes': {  2: {  'replication': {  2: <ProbDist 10.27-9.59=1.0>,
                                    3: <ProbDist 7.7-6.7=1.0>}},
               3: <ProbDist 8.6-8.3=1.0>,
               4: {  'replication': {  1: {  'colocated': {  0: <ProbDist 10.27-9.59=1.0>,
                                                  1: <ProbDist 9.59-8.6=1.0>}},
                                    2: <ProbDist 8.3-8.15=1.0>,
                                    3: <ProbDist 8.3-8.15=1.0>}},
               5: <ProbDist 9.59-8.6=1.0>,
               6: <ProbDist 8.6-8.3=1.0>,
               8: {  'replication': {  2: <ProbDist 10.6-10.27=1.0>,
                                    3: { 'blk_range': {  64: <ProbDist 10.27-9.59=1.0>,
                                                  128: <ProbDist 12.58-11.59=1.0>}}}},
               12: {  'replication': {  2: <ProbDist 13.9-12.58=1.0>,
                                    3: {  'data_size': {  1: <ProbDist 12.58-11.59=1.0>,
                                                  10: <ProbDist 26.58-23.39=1.0>}},
                                    6: <ProbDist 10.27-9.59=1.0>}},
               16: {  'replication': {  2: <ProbDist 12.58-11.59=1.0>,
                                    3: <ProbDist 11.59-11.04=1.0>,
                                    4: <ProbDist 10.27-9.59=1.0>}}}},
     18: {  'data_size': {  1: {  'nodes': {  1: <ProbDist 7.7-6.7=1.0>,
                                    2: {  'replication': {  2: <ProbDist 10.27-9.59=1.0>,
                                                  3: { 'blk_range': {  64: <ProbDist 7.7-6.7=1.0>,
                                                                128: <ProbDist 6.7-5.8=1.0>}}}},
                                    3: <ProbDist 7.7-6.7=1.0>,
                                    4: {  'replication': {  1: {  'colocated': {  0: <ProbDist 10.6-
10.27=1.0>,
                                                                1: <ProbDist 8.6-8.3=1.0>}},
                                                  2: <ProbDist 9.59-8.6=1.0>,
                                                  3: <ProbDist 9.59-8.6=1.0>,
                                                  4: { 'blk_range': {  64: <ProbDist 9.59-8.6=1.0>,
                                                                128: <ProbDist 8.3-8.15=1.0>}}}},
                                    5: <ProbDist 8.6-8.3=1.0>,
                                    6: <ProbDist 8.6-8.3=1.0>,
                                    8: {  'replication': {  2: <ProbDist 11.59-11.04=1.0>,
                                                  3: <ProbDist 9.59-8.6=1.0>}},
                                    12: {  'replication': {  2: <ProbDist 9.59-8.6=1.0>,
                                                  3: <ProbDist 13.9-12.58=1.0>,
                                                  5: <ProbDist 13.9-12.58=1.0>,
                                                  6: <ProbDist 9.59-8.6=1.0>}},
                                    16: {  'replication': {  2: <ProbDist 9.59-8.6=1.0>,
                                                  3: <ProbDist 14.56-13.9=1.0>,
                                                  4: <ProbDist 10.6-10.27=1.0>}}}},
               10: {  'nodes': {  2: <ProbDist 12.58-11.59=1.0>,
                          3: <ProbDist 8.3-8.15=1.0>,
                          4: { 'blk_range': {  64: <ProbDist 11.59-11.04=1.0>,
                                          128: <ProbDist 10.27-9.59=1.0>,
                                          256: <ProbDist 11.59-11.04=1.0>}},
                          5: <ProbDist 10.6-10.27=1.0>,
                          8: { 'blk_range': {  64: <ProbDist 10.27-9.59=1.0>,
```

```
                                              128: <ProbDist 10.6-10.27=1.0>}},
                             12: {   'replication': {   3: <ProbDist 14.56-13.9=1.0>,
                                                        4: <ProbDist 10.6-10.27=1.0>,
                                                        5: <ProbDist 11.59-11.04=1.0>}},
                             16: {   'replication': {   3: {   'colocated': {   1: <ProbDist 12.58-
11.59=0.5, 9.59-8.6=0.5>}}}}}}}},
        19: {   'data_size': {   1: {   'replication': {   1: {   'colocated': {   0: <ProbDist 30.38-26.58=1.0>,
                                                                                   1: <ProbDist 26.58-23.39=1.0>}},
                                                          2: {   'nodes': {   2: <ProbDist 42.27-37.19=1.0>,
                                                                              4: <ProbDist 23.39-21.5=1.0>,
                                                                              8: <ProbDist 37.19-30.38=1.0>,
                                                                              12: <ProbDist 30.38-26.58=1.0>,
                                                                              16: <ProbDist 30.38-26.58=1.0>}},
                                                          3: {   'nodes': {   2: {   'blk_range': {   64: <ProbDist 20.53-
19.57=1.0>,
                                                                                                     128: <ProbDist 18.56-17.58=1.0>}},
                                                                              3: <ProbDist 26.58-23.39=1.0>,
                                                                              4: <ProbDist 21.5-20.53=1.0>,
                                                                              5: <ProbDist 26.58-23.39=1.0>,
                                                                              6: <ProbDist 26.58-23.39=1.0>,
                                                                              8: <ProbDist 26.58-23.39=1.0>,
                                                                              12: <ProbDist 37.19-30.38=1.0>,
                                                                              16: <ProbDist 37.19-30.38=1.0>}},
                                                          4: <ProbDist 30.38-26.58=1.0>,
                                                          6: <ProbDist 26.58-23.39=1.0>}},
                                  10: <ProbDist 37.19-30.38=1.0>}},
        20: {   'nodes': {   2: {   'replication': {   2: <ProbDist 23.39-21.5=1.0>,
                                                       3: {   'blk_range': {   64: <ProbDist 9.59-8.6=1.0>,
                                                                               128: <ProbDist 8.15-7.7=1.0>}}}},
                             3: <ProbDist 10.27-9.59=1.0>,
                             4: {   'replication': {   1: {   'colocated': {   0: <ProbDist 12.58-11.59=1.0>,
                                                                               1: <ProbDist 11.59-11.04=1.0>}},
                                                       2: <ProbDist 11.04-10.6=1.0>,
                                                       3: <ProbDist 10.6-10.27=1.0>}},
                             5: <ProbDist 10.27-9.59=1.0>,
                             6: <ProbDist 10.6-10.27=1.0>,
                             8: {   'replication': {   2: <ProbDist 13.9-12.58=1.0>,
                                                       3: {   'blk_range': {   64: <ProbDist 12.58-11.59=1.0>,
                                                                               128: <ProbDist 10.6-10.27=1.0>}}}},
                             12: {   'replication': {   2: <ProbDist 12.58-11.59=1.0>,
                                                        3: <ProbDist 13.9-12.58=1.0>,
                                                        6: <ProbDist 10.6-10.27=1.0>}},
                             16: {   'replication': {   2: <ProbDist 11.59-11.04=1.0>,
                                                        3: <ProbDist 14.56-13.9=1.0>,
                                                        4: <ProbDist 11.59-11.04=1.0>}}}},
        21: {   'replication': {   1: <ProbDist 7.7-6.7=1.0>,
                                   2: {   'nodes': {   2: <ProbDist 8.3-8.15=1.0>,
                                                       4: <ProbDist 6.7-5.8=1.0>,
                                                       8: <ProbDist 8.6-8.3=1.0>,
                                                       12: <ProbDist 7.7-6.7=1.0>,
                                                       16: <ProbDist 7.7-6.7=1.0>}},
                                   3: {   'nodes': {   2: <ProbDist 5.8-4.54=1.0>,
                                                       3: <ProbDist 6.7-5.8=1.0>,
```

```
                              4: <ProbDist 6.7-5.8=1.0>,
                              5: <ProbDist 5.8-4.54=1.0>,
                              6: <ProbDist 6.7-5.8=1.0>,
                              8: <ProbDist 6.7-5.8=1.0>,
                             12: {   'data_size': {   1: <ProbDist 9.59-8.6=1.0>,
                                                     10: <ProbDist 20.53-19.57=1.0>}},
                             16: <ProbDist 8.6-8.3=1.0>}},
                      4: <ProbDist 7.7-6.7=1.0>,
                      6: <ProbDist 6.7-5.8=1.0>}},
      22: {   'data_size': {   1: {   'nodes': {   2: {   'replication': {   2: <ProbDist 4.54-3.9=1.0>,
                                            3: {   'blk_range': {   64: <ProbDist 3.3-2.54=1.0>,
                                                                   128: <ProbDist 2.54-1.52=1.0>}}}},
                              3: <ProbDist 3.3-2.54=1.0>,
                              4: {   'replication': {   1: {   'colocated': {   0: <ProbDist 4.54-
3.9=1.0>,
                                                                               1: <ProbDist 3.6-3.3=1.0>}},
                                                       2: <ProbDist 3.3-2.54=1.0>,
                                                       3: <ProbDist 3.3-2.54=1.0>}},
                              5: <ProbDist 3.9-3.6=1.0>,
                              6: <ProbDist 3.6-3.3=1.0>,
                              8: {   'replication': {   2: <ProbDist 4.54-3.9=1.0>,
                                                       3: {   'blk_range': {   64: <ProbDist 5.8-4.54=1.0>,
                                                                              128: <ProbDist 4.54-3.9=1.0>}}}},
                             12: {   'replication': {   2: <ProbDist 4.54-3.9=1.0>,
                                                       3: <ProbDist 5.8-4.54=1.0>,
                                                       6: <ProbDist 3.3-2.54=1.0>}},
                             16: {   'replication': {   2: <ProbDist 3.3-2.54=1.0>,
                                                       3: <ProbDist 4.54-3.9=1.0>,
                                                       4: <ProbDist 3.3-2.54=1.0>}}}},
                      10: <ProbDist 10.6-10.27=1.0>}},
      23: {   'nodes': {   2: {   'replication': {   2: <ProbDist 26.58-23.39=1.0>,
                                            3: {   'blk_range': {   64: <ProbDist 18.56-17.58=1.0>,
                                                                   128: <ProbDist 16.58-15.8=1.0>}}}},
                      3: <ProbDist 23.39-21.5=1.0>,
                      4: {   'replication': {   1: {   'colocated': {   0: <ProbDist 26.58-23.39=1.0>,
                                                                       1: <ProbDist 23.39-21.5=1.0>}},
                                               2: <ProbDist 19.57-18.56=1.0>,
                                               3: <ProbDist 19.57-18.56=1.0>}},
                      5: <ProbDist 21.5-20.53=1.0>,
                      6: <ProbDist 23.39-21.5=1.0>,
                      8: {   'blk_range': {   64: <ProbDist 26.58-23.39=1.0>,
                                             128: <ProbDist 21.5-20.53=1.0>}},
                     12: {   'replication': {   2: <ProbDist 26.58-23.39=1.0>,
                                               3: {   'data_size': {   1: <ProbDist 30.38-26.58=1.0>,
                                                                      10: <ProbDist 56.12-42.27=1.0>}},
                                               6: <ProbDist 26.58-23.39=1.0>}},
                     16: <ProbDist 26.58-23.39=1.0>}},
      24: {   'nodes': {   2: {   'replication': {   2: <ProbDist 17.58-16.58=1.0>,
                                            3: {   'blk_range': {   64: <ProbDist 12.58-11.59=1.0>,
                                                                   128: <ProbDist 11.59-11.04=1.0>}}}},
                      3: <ProbDist 14.56-13.9=1.0>,
                      4: {   'replication': {   1: {   'colocated': {   0: <ProbDist 17.58-16.58=1.0>,
                                                                       1: <ProbDist 16.58-15.8=1.0>}},
```

```
                                2: <ProbDist 13.9-12.58=1.0>,
                                3: <ProbDist 13.9-12.58=1.0>}},
                    5: <ProbDist 14.56-13.9=1.0>,
                    6: <ProbDist 14.56-13.9=1.0>,
                    8: {   'replication': {   2: <ProbDist 18.56-17.58=1.0>,
                                              3: <ProbDist 15.8-14.56=1.0>}},
                    12: {   'replication': {   2: <ProbDist 16.58-15.8=1.0>,
                                               3: {   'data_size': {   1: <ProbDist 21.5-20.53=1.0>,
                                                                       10: <ProbDist 17.58-16.58=1.0>}},
                                               6: <ProbDist 16.58-15.8=1.0>}},
                    16: {   'replication': {   2: <ProbDist 15.8-14.56=1.0>,
                                               3: <ProbDist 19.57-18.56=1.0>,
                                               4: <ProbDist 21.5-20.53=1.0>}}}},
     25: {   'replication': {   1: <ProbDist 17.58-16.58=1.0>,
                                2: {   'nodes': {   2: <ProbDist 18.56-17.58=1.0>,
                                                    4: <ProbDist 14.56-13.9=1.0>,
                                                    8: <ProbDist 18.56-17.58=1.0>,
                                                    12: <ProbDist 17.58-16.58=1.0>,
                                                    16: <ProbDist 16.58-15.8=1.0>}},
                                3: {   'nodes': {   2: {   'blk_range': {   64: <ProbDist 12.58-11.59=1.0>,
                                                                            128: <ProbDist 11.59-11.04=1.0>}},
                                                    3: <ProbDist 15.8-14.56=1.0>,
                                                    4: <ProbDist 13.9-12.58=1.0>,
                                                    5: <ProbDist 14.56-13.9=1.0>,
                                                    6: <ProbDist 15.8-14.56=1.0>,
                                                    8: {   'blk_range': {   64: <ProbDist 15.8-14.56=1.0>,
                                                                            128: <ProbDist 16.58-15.8=1.0>}},
                                                    12: {   'data_size': {   1: <ProbDist 21.5-20.53=1.0>,
                                                                             10: <ProbDist 20.53-19.57=1.0>}},
                                                    16: <ProbDist 20.53-19.57=1.0>}},
                                4: <ProbDist 17.58-16.58=1.0>,
                                6: <ProbDist 17.58-16.58=1.0>}},
     26: {   'replication': {   1: <ProbDist 12.58-11.59=1.0>,
                                2: {   'nodes': {   2: <ProbDist 12.58-11.59=1.0>,
                                                    4: <ProbDist 9.59-8.6=1.0>,
                                                    8: <ProbDist 13.9-12.58=1.0>,
                                                    12: <ProbDist 11.04-10.6=1.0>,
                                                    16: <ProbDist 11.59-11.04=1.0>}},
                                3: {   'nodes': {   2: {   'blk_range': {   64: <ProbDist 8.6-8.3=1.0>,
                                                                            128: <ProbDist 8.3-8.15=1.0>}},
                                                    3: <ProbDist 10.6-10.27=1.0>,
                                                    4: <ProbDist 10.6-10.27=1.0>,
                                                    5: <ProbDist 9.59-8.6=1.0>,
                                                    6: <ProbDist 10.6-10.27=1.0>,
                                                    8: <ProbDist 10.6-10.27=1.0>,
                                                    12: {   'data_size': {   1: <ProbDist 15.8-14.56=1.0>,
                                                                             10: <ProbDist 12.58-11.59=1.0>}},
                                                    16: <ProbDist 13.9-12.58=1.0>}},
                                4: <ProbDist 11.59-11.04=1.0>,
                                6: <ProbDist 10.6-10.27=1.0>}},
     27: {   'colocated': {   0: <ProbDist 3.3-2.54=1.0>,
                              1: {   'replication': {   1: <ProbDist 2.54-1.52=1.0>,
                                                        2: {   'nodes': {   2: <ProbDist 3.9-3.6=1.0>,
```

```
                                    4: <ProbDist 2.54-1.52=1.0>,
                                    8: <ProbDist 3.3-2.54=1.0>,
                                   12: <ProbDist 2.54-1.52=1.0>,
                                   16: <ProbDist 2.54-1.52=1.0>}},
                        3: {   'nodes': {   2: <ProbDist 2.54-1.52=1.0>,
                                    3: <ProbDist 2.54-1.52=1.0>,
                                    4: <ProbDist 2.54-1.52=1.0>,
                                    5: <ProbDist 2.54-1.52=1.0>,
                                    6: <ProbDist 2.54-1.52=1.0>,
                                    8: <ProbDist 2.54-1.52=1.0>,
                                   12: {   'data_size': {   1: <ProbDist 3.6-3.3=1.0>,
                                                   10: <ProbDist 2.54-1.52=1.0>}},
                                   16: <ProbDist 3.3-2.54=1.0>}},
                        4: <ProbDist 3.6-3.3=1.0>,
                        6: <ProbDist 2.54-1.52=1.0>}}},
         28: {   'nodes': {   2: {   'replication': {   2: <ProbDist 71.12-58.39=1.0>,
                                    3: {   'blk_range': {   64: <ProbDist 56.12-42.27=1.0>,
                                                  128: <ProbDist 42.27-37.19=1.0>}}}},
                        3: <ProbDist 56.12-42.27=1.0>,
                        4: <ProbDist 56.12-42.27=1.0>,
                        5: <ProbDist 56.12-42.27=1.0>,
                        6: <ProbDist 56.12-42.27=1.0>,
                        8: {   'replication': {   2: <ProbDist 71.12-58.39=1.0>,
                                    3: <ProbDist 56.12-42.27=1.0>}},
                        12: {   'replication': {   2: <ProbDist 58.39-56.12=1.0>,
                                    3: <ProbDist 71.12-58.39=1.0>,
                                    6: <ProbDist 56.12-42.27=1.0>}},
                        16: {   'replication': {   2: <ProbDist 56.12-42.27=1.0>,
                                    3: <ProbDist 71.12-58.39=1.0>,
                                    4: <ProbDist 71.12-58.39=1.0>}}}},
         29: {   'nodes': {   2: {   'replication': {   2: <ProbDist 8.6-8.3=1.0>,
                                    3: {   'blk_range': {   64: <ProbDist 6.7-5.8=1.0>,
                                                  128: <ProbDist 5.8-4.54=1.0>}}}},
                        3: <ProbDist 6.7-5.8=1.0>,
                        4: {   'replication': {   1: {   'colocated': {   0: <ProbDist 8.6-8.3=1.0>,
                                                            1: <ProbDist 7.7-6.7=1.0>}},
                                    2: <ProbDist 6.7-5.8=1.0>,
                                    3: <ProbDist 6.7-5.8=1.0>}},
                        5: <ProbDist 7.7-6.7=1.0>,
                        6: <ProbDist 6.7-5.8=1.0>,
                        8: {   'replication': {   2: <ProbDist 9.59-8.6=1.0>,
                                    3: {   'blk_range': {   64: <ProbDist 8.15-7.7=1.0>,
                                                  128: <ProbDist 7.7-6.7=1.0>}}}},
                        12: {   'replication': {   2: <ProbDist 9.59-8.6=1.0>,
                                    3: {   'data_size': {   1: <ProbDist 8.6-8.3=1.0>,
                                                   10: <ProbDist 9.59-8.6=1.0>}},
                                    6: <ProbDist 7.7-6.7=1.0>}},
                        16: {   'replication': {   2: <ProbDist 8.3-8.15=1.0>,
                                    3: <ProbDist 9.59-8.6=1.0>,
                                    4: <ProbDist 9.59-8.6=1.0>}}}},
         30: {   'nodes': {   2: {   'replication': {   2: <ProbDist 11.59-11.04=1.0>,
                                    3: {   'blk_range': {   64: <ProbDist 7.7-6.7=1.0>,
                                                  128: <ProbDist 8.6-8.3=1.0>}}}},
```

```
                    3: <ProbDist 9.59-8.6=1.0>,
                 4: {   'replication': {   1: {   'colocated': {   0: <ProbDist 8.6-8.3=1.0>,
                                                                   1: <ProbDist 9.59-8.6=1.0>}},
                                           2: <ProbDist 8.3-8.15=1.0>,
                                           3: <ProbDist 7.7-6.7=1.0>}},
                 5: <ProbDist 8.15-7.7=1.0>,
                 6: <ProbDist 9.59-8.6=1.0>,
                 8: {   'blk_range': {   64: <ProbDist 9.59-8.6=1.0>,
                                         128: <ProbDist 8.6-8.3=1.0>}},
                 12: {   'replication': {   3: <ProbDist 10.6-10.27=1.0>,
                                            6: <ProbDist 9.59-8.6=1.0>}},
                 16: {   'replication': {   2: <ProbDist 8.3-8.15=1.0>,
                                            3: <ProbDist 10.27-9.59=1.0>,
                                            4: <ProbDist 13.9-12.58=1.0>}}}}}}
```